\documentstyle[sprocl,psfig,here]{article}

\def\be{\begin{equation}}
\def\ee{\end{equation}}
\def\bea{\begin{eqnarray}}
\def\eea{\end{eqnarray}}

\begin{document}
\begin{flushright}
HUPD-9808 \\
hep-ph/9803448 \\
\end{flushright}
\title{THE SMALL $x$ BEHAVIOR OF $g_1$ IN THE RESUMMED APPROACH 
\footnote{Invited talk presented by
H.Tochimura at the International Symposium on
QCD Corrections and New Physics, HIROSHIMA, 1997.}}
\author{YUICHIRO KIYO, JIRO KODAIRA and HIROSHI TOCHIMURA}
\address{Dept. of Physics, Hiroshima University Higashi-Hiroshima
739-8526, JAPAN}
%%%%%%%%%%%%%%%%%%%%%%%%%%%%%%%%%%%%%%%%%%%%%%%%%%%%%%%%%%%%%%
% You may repeat \author \address as often as necessary      %
%%%%%%%%%%%%%%%%%%%%%%%%%%%%%%%%%%%%%%%%%%%%%%%%%%%%%%%%%%%%%%
\maketitle
\abstracts{
\hspace*{1ex}
The double logarithmic terms $\alpha_{s} \ln^{2}x $ are important
to predict precisely
the small $x$ behavior of the spin
structure function $g_{1}$.
We numerically analyze the evolution of the flavor non-singlet
$g_{1}$ including the
all-order resummed effect of these terms.
It is pointed out that the next-to-leading logarithmic
corrections
produce an unexpectedly large suppression factor over the
experimentally accessible range of $x$ and $Q^{2}$.
This implies that the next-to-leading logarithmic
contributions are very important in order to obtain
a definite prediction.}
\section{Introduction}
Recently many experimental and theoretical works have 
been devoted to the polarized structure function $g_1$.
Especially, it is very important and desirable to know
the small $x$ behavior of $g_1$ in the light of the Bjorken
and Ellis-Jaffe sum rules. 
Since the verification of these sum rules
requires the knowledge of the structure function over the entire
$x$ regions,
one has to rely on the theoretical prediction
in the experimentally inaccessible small $x$ region.
Although the Regge prediction has usually been assumed 
for the extrapolation of the experimental data
to the small $x$ region, the recent data show 
a clear departure from this naive Regge prediction ~\cite{smc}.
This fact means that the perturbative QCD effects are very important.\\
\hspace*{1ex}
Recently, various extrapolations have been proposed 
using the DGLAP equation ~\cite{grsv}.
However,it has been known that there appear the double
logarithmic terms $\alpha_s ln^2 x$ in the perturbative calculations. 
These logarithmic corrections seem to give large effects in the small
$x$ region and are important
to get more reliable predictions on the small $x$ 
behavior of $g_1$. 
Bartels, Ermolaev and Ryskin~\cite{bartels} have given
the resummed expression for 
the partonic structure function  $g^{parton}_{1}$ 
by using the Infra-Red Evolution Equation and confirmed the old
result by Kirschner and Lipatov~\cite{kili}.
They claim that the resummed effects may be important.
But, when extracting the physical structure
function of hadrons from the partonic one, 
there is possibility that 
their conclusion at the parton level is not necessarily true.
Indeed,the recent numerical analysis by Bl\"umlein and
Vogt~\cite{blvo} shows
that there are no significant contributions from the resummation
of the leading logarithmic (LL) corrections in the HERA kinematical
region. 
The different conclusions between at the partonic and hadronic level might
come from the fact that the resummed \lq\lq coefficient function\rq\rq \ 
was not included in Ref.~\cite{blvo} because it
falls in the next-to-leading logarithmic (NLL) corrections
and depends on the factorization scheme adopted. 
It is also to be noted that the slightly steep input
density was used in Ref.~\cite{blvo}.
The evolution, in general, strongly
depends on the input parton densities.
If one chooses a steep input function, the perturbative
contribution will be completely washed away.
So it will be also interesting to see the sensitivity of the
results to the choice of the input densities.\\
\hspace*{1ex}
In this report, we reanalyze the numerical impact  
of the resummed effects on the small $x$
behavior of $g_{1}$ (non-singlet part).
The coefficient function can not
be included consistently at present since the anomalous
dimension has been calculated only to the LL order.
However we consider the effects of this part
because we could firstly clarify an origin of the
above different conclusions and secondly get some
idea about the magnitude of the NLL order
corrections in the resummation approach. 
We also consider two different input densities: one is flat
corresponding to the naive Regge prediction and
the other is steep in the small $x$ region. Details of the
calculations may be found in Ref.~\cite{kkt}. 
%
%---------- Section 2 ----------------------------------------
\section{Resummed structure function $g_1$ } 
The flavor non-singlet part of $g_{1}$ in the moment space is given by,
\[ g_{1}(Q^{2},N) \equiv \int^{1}_{0}dx x^{N-1} g_{1}(Q^{2},x)
  = \frac{\langle e^{2} \rangle}{2}
     C (\alpha_{s}(Q^{2}),N) \Delta q (Q^{2},N)\ ,\]
where
\[ C (\alpha_{s}(Q^{2}),N) \equiv \int^{1}_{0} d x x^{N-1}
    C (\alpha_s (Q^{2}),x) \ \ ,\ \ \Delta q (Q^{2},N)
     \equiv \int^{1}_{0} d x x^{N-1} \Delta q (Q^{2},x)\ ,\]
and $\Delta q (Q^{2},x)$ ($C$) is the flavor non-singlet
combination of the polarized parton densities (the coefficient function).
$\langle e^{2} \rangle$ is the average of the quark's electric charge. 
The DGLAP equation is,
\be
  Q^{2} \frac{\partial}{\partial Q^{2}} \Delta q (Q^{2},N) 
   = - \gamma (\alpha_{s}(Q^{2}),N)
   \Delta q (Q^{2},N) \ .\label{eqn:rgeq} 
\ee
Here the anomalous dimension $\gamma$ is the moment of the
\lq\lq splitting\rq\rq\ function.
The coefficient function $C (\alpha_{s},N)$ and the
anomalous dimension $\gamma (\alpha_{s},N)$ can be calculated
perturbatively and are 
 expanded
in the powers of $\alpha_s$,
\[ C (\alpha_{s},N) = 1 + \sum_{k=1}^{\infty}
     c^{k}(N) \bar{\alpha}_{s}^{k} \ \ ,\ \ \gamma (\alpha_{s},N)
   = \sum_{k=1}^{\infty} \gamma^{k}(N) \bar{\alpha}_{s}^{k} \ .\]
%1
where $\bar{\alpha}_{s} \equiv \frac{\alpha_s}{4\pi}$ .
%The important terms of the coefficient function $C$ 
%and the anomalous dimension $\gamma $ in the small $x$  
%is the pole singular term at $N=0$  since the following relation
%exists,
%\[
%\frac{1}{N^{k}} = \frac{(-1)^k}{k!}\int^1_0dx x^{k-1}ln^{k}x .
%\]
%
The singular behaviors of the anomalous dimension and the coefficient 
function as $x \to 0$ appear as the pole singularities at $N=0$.
We must resum these singular terms to all orders to
get a reliable prediction. 
This resummation has been done in Refs.~\cite{bartels}~\cite{kili} and 
the resummed part of these functions are given 
as follows,
\bea
   \hat{\gamma}(\alpha_{s},N) 
    &\equiv& \lim_{N \to 0} \gamma(\alpha_{s},N) = -
   f^{-}_{0}(N)/8 \pi^{2} \ ,\label{resuma}\\
  \hat{C}(\alpha_{s},N) &\equiv& \lim_{N \to 0} C(\alpha_{s},N)
     = \frac{N}{N-f^{(-)}_{0}(N) / 8\pi^{2}} \ .\label{resumc}
\eea
Here $f_{0}^{-}$ is given by,
\[   f^{-}_{0}(N) = 4 \pi^{2}N
  \left( 1-\sqrt{1- 8 C_F \frac{\bar{\alpha}_s}{N^{2}}
  \left[ 1-\frac{1}{2\pi^{2}N}f^{+}_{8}(N) \right]} \right) \ .\]
with
\[ f^{+}_{8} (N) = 16 \pi^2 N_{c} \bar{\alpha}_s \frac{d}{dN}
   \ln(e^{z^{2}/4}D_{-1/2N_{c}^{2}}(z)) \quad , \quad
           z = \frac{N}{\sqrt{2 N_{c} \bar{\alpha}_s}}\ .\]
$ D_{p}(z)$ is the parabolic cylinder function.
%and 
%$\bar{\alpha}_s\equiv \frac{\alpha_s}{4\pi}$.
%
One can easily see by expanding Eqs.(\ref{resuma},\ref{resumc}) 
in terms of $\alpha_{s}$ that 
the resummed expressions
Eqs.(\ref{resuma},\ref{resumc}) reproduce
the known NLO results in the $\overline{\rm MS}$
scheme. 
Therefore, it is quite plausible that
Eqs.(\ref{resuma},\ref{resumc}) correctly sum up
the \lq\lq leading\rq\rq\ singularities to all orders.\\
\hspace*{1ex}
Before going to the numerical analysis,
we explain the reason why the resummed coefficient part was discarded
in the analysis of Bl\"umelein and Vogt.
For definiteness, let us use 
the so-called DIS scheme.
The parton density and the anomalous dimension in the DIS scheme
are obtained by making the transformations,
\be \Delta q
 \rightarrow  \Delta q^{DIS} \equiv C \Delta q \ \ ,\ \ \gamma^{DIS}
     \equiv C \gamma C^{-1} - \beta(\alpha_{s})
    \frac{\partial}{\partial \alpha_{s}} \ln C \ .
\label{eqn:dis}
\ee
Using the resummed $\hat{\gamma}$ and $\hat{C}$ 
Eqs.(\ref{resuma},\ref{resumc}),
we get the anomalous dimension in the DIS scheme,
\be
 \hat{\gamma}^{DIS} = N \sum^{\infty}_{k=1}\hat{\gamma}^{k}
      \left( \frac{\bar{\alpha}_{s}}{N^{2}} \right)^{k}
  + \beta_{0} N^2 \sum^{\infty}_{k=2}
    \hat{d}^{k}
    \left( \frac{\bar{\alpha}_{s}}{N^{2}} \right)^{k} 
  + {\cal O} \left( N^3 
      \left( \frac{\bar{\alpha}_{s}}{N^{2}} \right)^{k} \right) \ ,
     \label{resumdis}
\ee
where the second term comes from the resummed coefficient function
and $\hat{d}^k$ are numerical numbers independent of $N$.
Since the second term has an extra $N$  in comparison with
the first term in the each power of $\alpha_s$,
the resummed coefficient function
belongs to the NLL order corrections 
in the context of the resummation approach.
This is the reason why the LL analysis of Bl\"umlein and Vogt
dropped the resummed coefficient function.
One must include the NLL order anomalous
dimension, which has not yet been available,
to perform a consistent analysis at the NLL level.
%---------- Section 3 ----------------------------------------
%
\section{Numerical consideration}
We show our numerical results.
From Eq.(\ref{eqn:dis}),
in DIS scheme, $g_1$ is given by
\be
g_1(x,Q^2) = \frac{\langle e^{2} \rangle}{2} \int_{ c-i \infty}^{c+i \infty}
\frac{d N}{2 \pi i} x^{- N} 
\Delta q^{DIS}(N,Q^2)
\label{eqn:g1-DIS}
\ee
The DGLAP equation Eq.(\ref{eqn:rgeq}) 
is easily solved with anomalous dimension
$\gamma^{DIS}$,
\[
\Delta q^{DIS}(N,Q^2) =  \exp
\left(-\int_{ \alpha_{s}(Q_{0}^{2})}^{\alpha_{s}(Q^{2})}
       \frac{d\alpha_{s}}{\beta} \gamma^{DIS} \right)\Delta q^{DIS}(N,Q_0^2).
\]
The anomalous dimension $\gamma^{DIS}$ 
which includes the resummed effects is obtained by using
Eqs.(\ref{resuma},\ref{resumc} ) and 
the relation Eq.(\ref{eqn:dis}),
\be
\gamma^{DIS}(N) = \bar{\alpha}_{s} \gamma^{1}(N)
   + \bar{\alpha}_{s}^{2} \gamma^{2}(N) + K (N,\alpha_{s})
  - \beta \frac{\partial}{\partial \alpha_{s}}
   \ln \left( 1 + \bar{\alpha}_{s} c^{1} + H(N,\alpha_{s}) \right) \ ,
\label{eqn:anoma}
\ee
where $\gamma^{1,2}$ and $c^{1}$ are respectively the exact anomalous
dimension and coefficient function at the one and two-loop fixed order
perturbation theory.
$K(N,\alpha_{s})$ ($H(N,\alpha_{s})$) is the resummed anomalous
dimension  Eq.(\ref{resuma}) 
(the coefficient function Eq.(\ref{resumc})) with $k = 1,2$ ($k = 0,1$)
terms being subtracted to avoid the double counting.
Although the anomalous dimension at $N=1$
should vanish due to the conservation of the (non-singlet) axial
vector current, 
the resummation of the leading singularities in $N$ does not respect
this symmetry.
One of prescriptions \footnote{Our final conclusion remains the same qualitatively
if we choose other prescription.}
to restore this symmetry is~\cite{blvo},
\[  K(N,\alpha_{s}) \to K(N,\alpha_{s}) (1-N) \ .\]
In order to estimate $g_1$, we must assume the appropriate function for
the input 
density $\Delta q^{DIS}(N,Q_0^2)$.
The explicit parameterization we use is~\cite{grsv},
\[ \Delta q^{DIS} (Q_0^{2},x) = N (\alpha , \beta , a )
       \eta x^{\alpha} (1 - x)^{\beta} ( 1 + a x)\ ,\]
where
$N$ is a normalization factor such that
$\int dx N x^{\alpha} (1 - x)^{\beta} ( 1 + a x) = 1$
and $\eta = \frac{1}{6} g_A $ ($g_A = 1.26 $) in
accordance with the  Bjorken sum rule and  we choose
the input scale $Q_{0}^{2} = 4 GeV^{2}$. 
Note that the small $x$ behavior is controlled by the parameter $\alpha$. 
Since we are also interested in 
the sensitivity of the final results to the small $x$ behavior of
input densities, 
we  choose two types of the input densities A and B:
A is a function which is flat at small $x$ ($x^{\alpha},\alpha \sim 0$),
and B is slightly steep ($\alpha \sim - 0.2$) .
A and B correspond to the following values of parameters,
\[  A \ (B)\ :\  \alpha = + 0.0 \ (- 0.2)\  , \ \beta = 3.09 \ (3.15)\  , \
    a = 2.23 \ (2.72)\ .\]
We put the flavor number 
$n_{f} = 4 $ and $\Lambda_{QCD}=0.23 GeV$.\\
\hspace*{1ex}
Now, let us explain how to perform
the Mellin inversion Eq.(\ref{eqn:g1-DIS})
which is an integral in the complex $N$-plane.
The contour integration along the imaginary axis
from $c-i \infty$ to $c + i \infty$
is numerically inconvenient due to the slow convergence of the integral
in the large $|N|$ region.
To get rid of this problem, we deformed the contour to the
line which have an angle $\phi$ ($\phi > \pi / 2$) from the real
$N$ axis. By this change of the contour, we have a damping factor
$\exp (|N| \ln (1/x) \cos\phi )$ which 
strongly suppresses the contribution from the large $|N|$ region.
In the integration along this new contour, we will be able to cut the
large $|N|$ region. Finally we have checked the stability of results
by changing the contour parameter.
One can find the details of this technique in Ref.~\cite{Reya}.\\
\hspace*{1ex}
First we estimate the case which includes only the LL correction.
The evolution kernel in this case is obtained by dropping   
$H(N,\alpha_{s})$ in Eq.(\ref{eqn:anoma}).
Fig.1a (1b) shows the LL results (dashed curves) after evolving
to $Q^{2} = 10, 10^2 ,$ $10^4 GeV^{2}$ from the A (B) input density
(dot-dashed line). The solid curves are the predictions of
the NLO-DGLAP evolution. 
These results show a tiny enhancement compared with the NLO-DGLAP
analysis and are consistent with those in Ref.~\cite{blvo}
The enhancement is, as expected, bigger when the input density is
flatter. However any significant differences are not seen between the
results from different input densities.
%\vspace{-10mm}
%%%%%%%%%%% Fig.1 %%%%%%%%%%%%%%%%%%%%%%%%%%%%%%%%%%%%%%%%
\begin{figure}[h]
\begin{center}
\begin{tabular}{cc}
\leavevmode\psfig{file=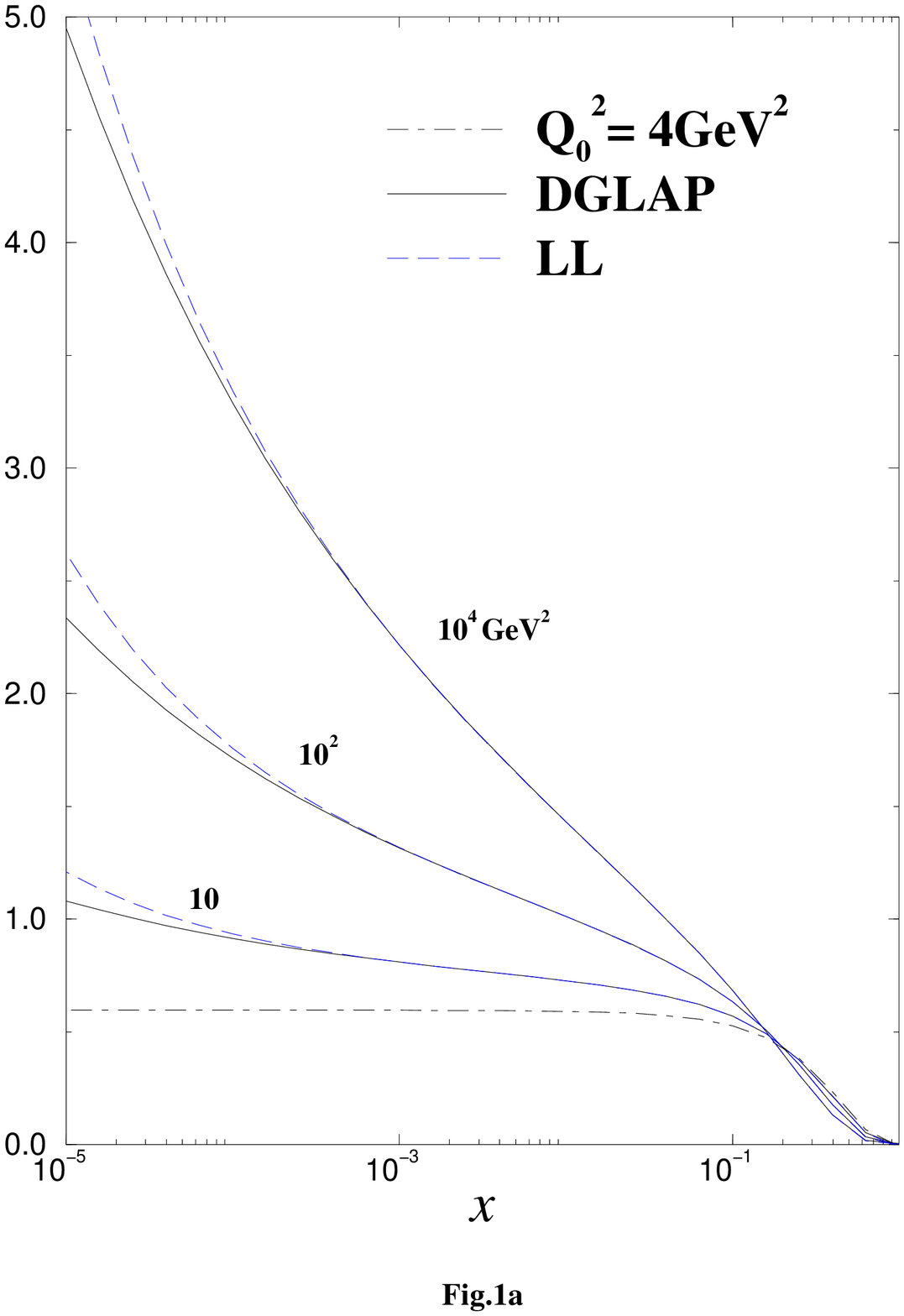,width=5.8 cm} &
\leavevmode\psfig{file=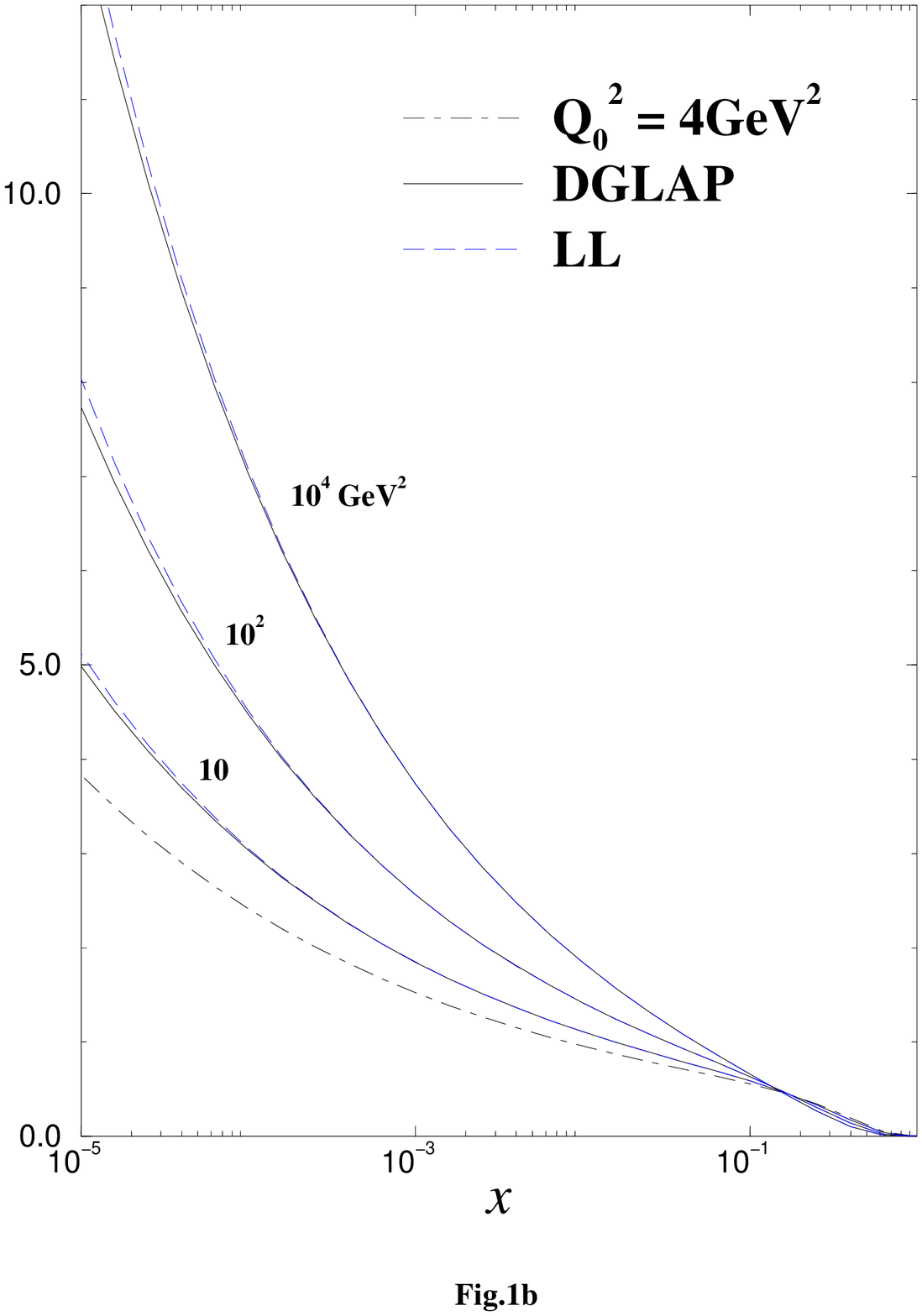,width=5.8 cm} 
\end{tabular}
\caption{The LL evolution as compared to the DGLAP results with the flat
input A (1a) and steep one B (1b).}
\end{center}
\end{figure}
%\vspace{-3mm}
Next, we include the NLL corrections from the resummed
\lq\lq coefficient function\rq\rq .
We show the results in Fig.2 by the dashed curves.
The results are rather surprising. The inclusion of
the coefficient function leads to a strong suppression
on the evolution of the structure function at small $x$.
%\vspace{-10mm}
\begin{figure}[h]
\begin{center}
\begin{tabular}{cc}
\leavevmode\psfig{file=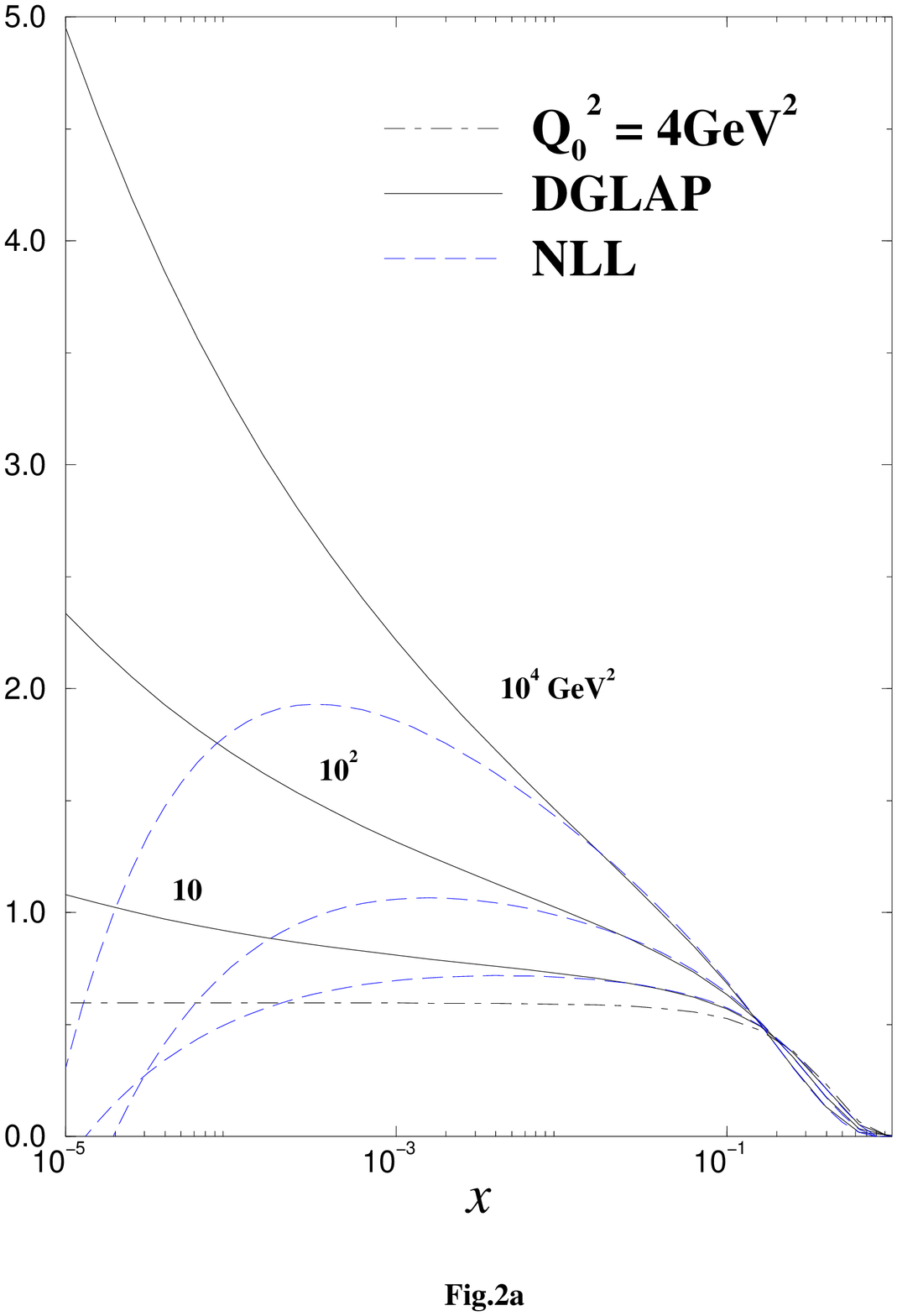,width=5.8 cm} &
\leavevmode\psfig{file=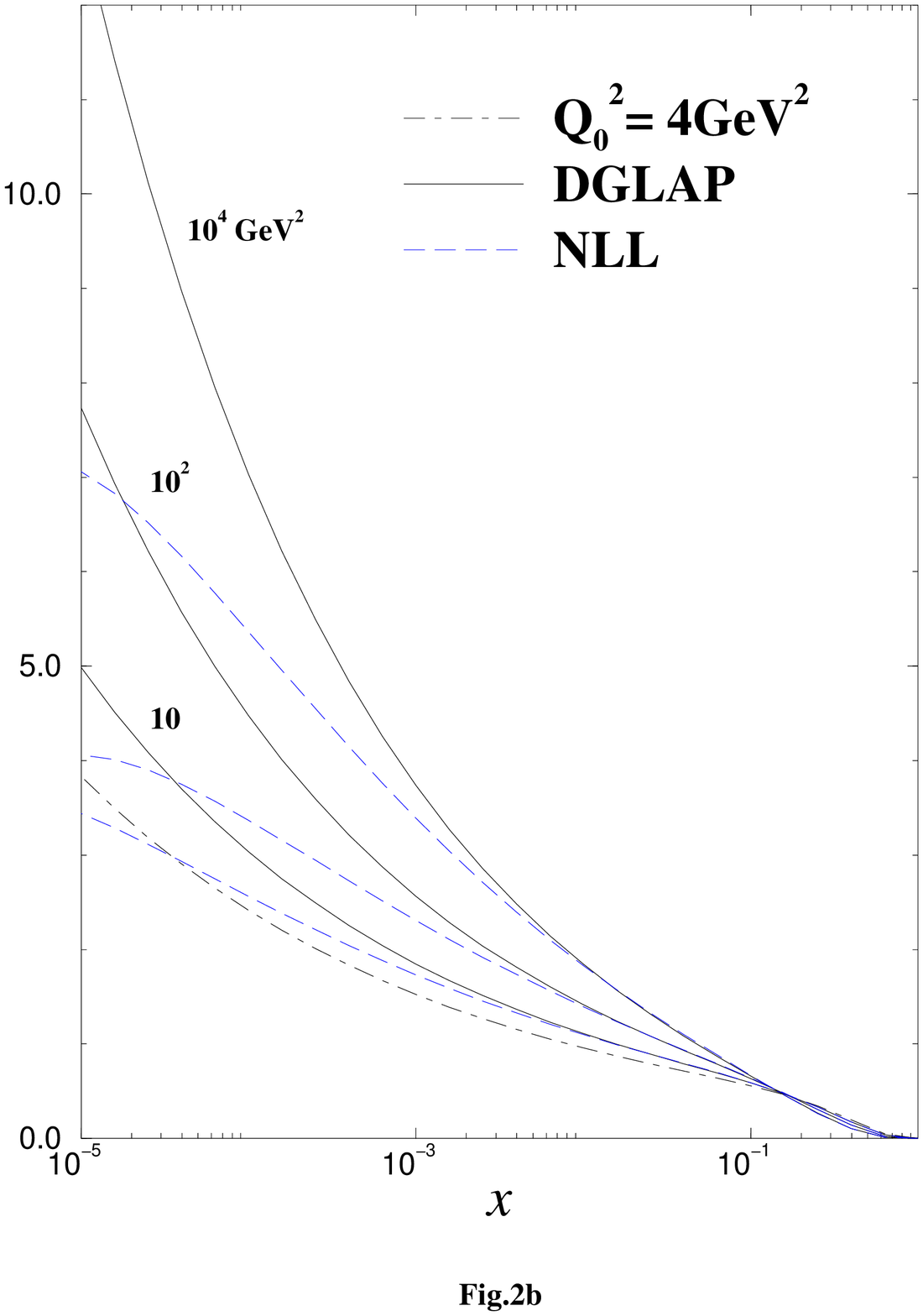,width=5.8 cm} 
\end{tabular}
\caption{The NLL evolution as compared to the DGLAP results with the
flat input A (2a) and steep one B (2b). }
\end{center}
\vspace{-3mm}
\end{figure}
%%%%%%%%%%%%%%%%%%%%%%%%%%%%%%%%%%%%%%%%%%%%%%%%%%%
%
To understand our numerical results, it will be helpful to remember the
perturbative expansion of the resummed results
Eqs.(\ref{resuma},\ref{resumc}).
Using the explicit values $N_C = 3 , C_F = 4/3$, we obtain
for the anomalous dimension in the DIS scheme
Eq.(\ref{resumdis}),
\bea
 \hat{\gamma}^{DIS} &=& N \left[ - 0.212
       \left( \frac{\alpha_{s}}{N^{2}} \right) \right. \nonumber\\
    & & \qquad - \, 
       0.068 \left. \left( \frac{\alpha_{s}}{N^{2}} \right)^{2} -
       0.017 \left( \frac{\alpha_{s}}{N^{2}} \right)^{3} -
       0.029 \left( \frac{\alpha_{s}}{N^{2}} \right)^{4} 
           + \cdots \right] \nonumber\\
    &+&  \, N^2 \left[
       0.141 \left( \frac{\alpha_{s}}{N^{2}} \right)^{2} +
       0.119 \left( \frac{\alpha_{s}}{N^{2}} \right)^{3} +
       0.069 \left( \frac{\alpha_{s}}{N^{2}} \right)^{4} 
           + \cdots \right] \label{numbergdis}\\ 
    &+&  \,\,\,  \cdots \ \ .\nonumber
\eea
Here note that: (1) the perturbative coefficients of the LL terms
(the first part of Eq.(\ref{numbergdis})) are negative
and those of the higher orders are rather small number.
This implies that the LL corrections push up
the structure function compared to the fixed-order DGLAP
evolution, but the deviations are expected to be small. 
(2) the perturbative ones from the NLL terms (the second part of
Eq.(\ref{numbergdis})), however, are positive and
somehow large compared with those of the LL terms.
This positivity of the NLL terms has the effect of
decreasing the structure function. 
It might be also helpful to {\sl assume} that the saddle-point dominates
the Mellin inversion Eq.(\ref{eqn:g1-DIS}). We have numerically estimated
the approximate position of the
saddle-point and found that the saddle-point stays around
$N_{\rm SP} \sim 0.31$ in the region of $x \sim 10^{-5}$ to $10^{-2}$.
By looking at the explicit values of the coefficients in
Eq.(\ref{numbergdis}), the position of the saddle-point seems to suggest that
the NLL terms can not be neglected. Since the coefficients from the higher
order terms are not so large numerically, it is also
expected that the terms  which
lead to sizable effects on the evolution may be only first few terms in the
perturbative series in the region of $x$ we are interested in.
%\vspace{-10mm}
%%%%%%%%%%%%%%%%% Fig.3 %%%%%%%%%%%%%%%%
\begin{figure}[h]
\begin{center}
\begin{tabular}{cc}
\leavevmode\psfig{file=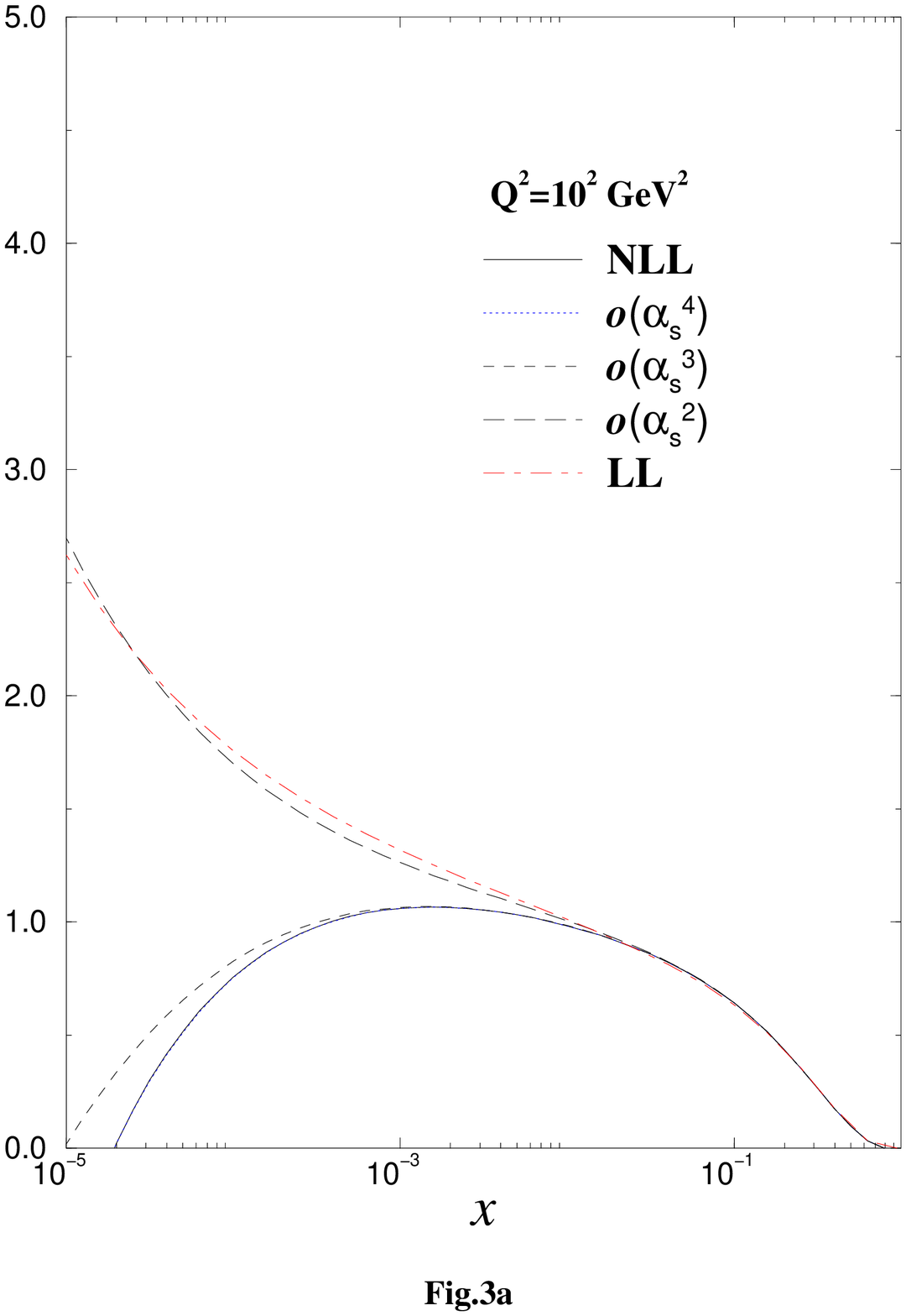,width=5.8cm} &
\leavevmode\psfig{file=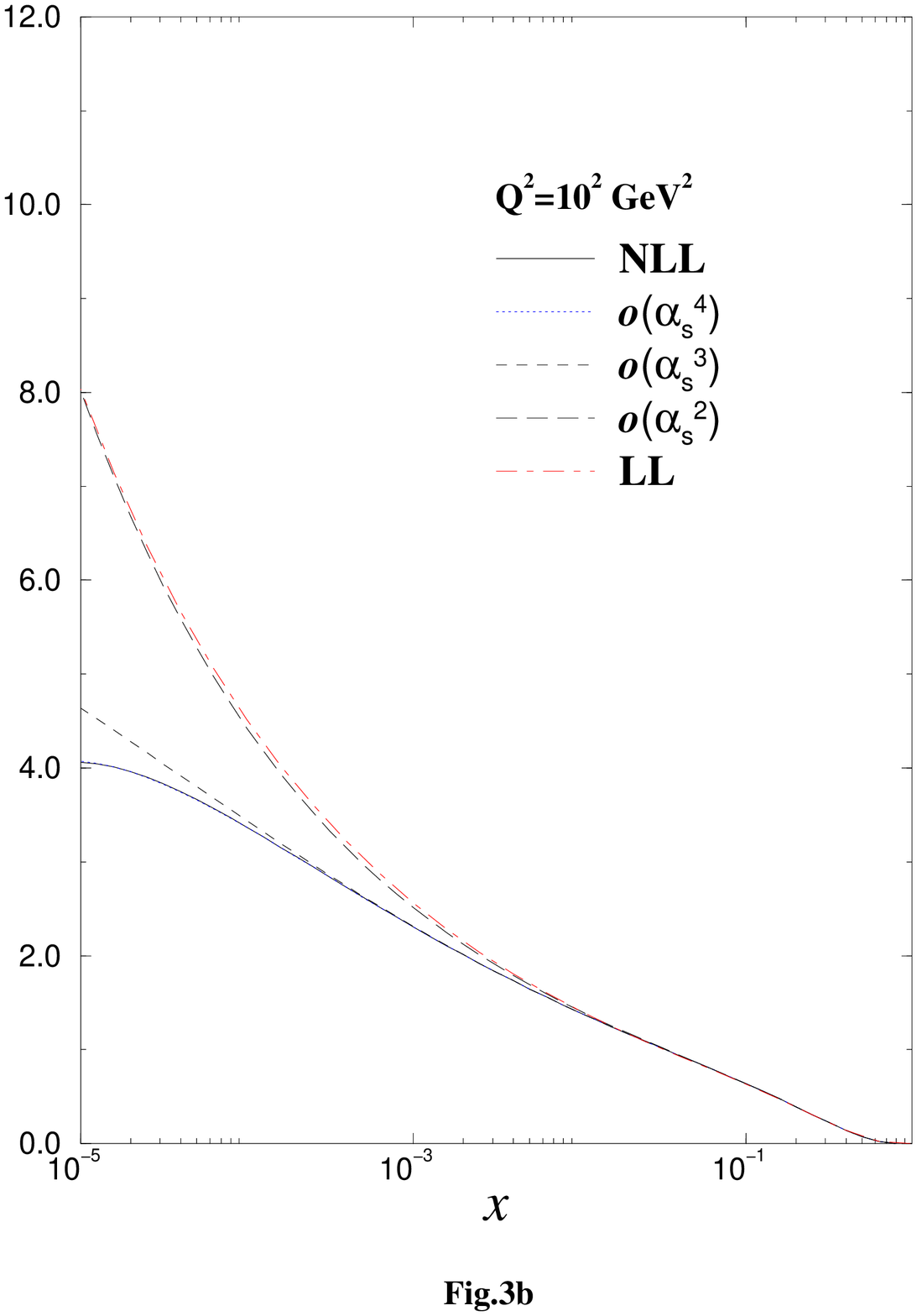,width=5.8cm} 
\end{tabular}
\caption{Contributions from the fixed order terms in the NLL
resummation with the flat input A (3a) and steep one B
(3b).}
\end{center}
\end{figure}
%\vspace{-3mm}
Fig.3a (3b) shows the numerical results of the
contribution from each terms of the NLL corrections in
Eq.(\ref{numbergdis}) at $Q^{2} = 10^2 GeV^{2}$ with the A (B)
type input density.
The solid (dot-dashed) line corresponds to the NLL (LL) result.
The long-dashed, dashed and dotted lines correspond
respectively to the case in which
the terms up to the order $\alpha_{s}^2$, $\alpha_{s}^3$, $\alpha_{s}^4$,
are kept in the NLL contributions. One can see that the
dotted line already coincides with the full NLL (solid) line.
These considerations could help us to understand why the NLL corrections
turns out to give large effects.
\section{summary}
We have numerically studied the small $x$ behavior of
the flavor non-singlet $g_{1}$ 
by taking into account the resummed effect of $\alpha_s ln^2 x$.
Our LL analysis is consistent with the results by 
Bl\"umlein and Vogt~\cite{blvo}.
We have also performed the analysis which includes a part of the NLL
corrections from the
resummed  coefficient function in the light of the assertion
of Bartels, Ermolaev and
Ryskin~\cite{bartels}, though this is not theoretically consistent.
Our results suggest that the LL analysis is unstable
in the sense that a large suppression effect comes from the resummed
coefficient function which should be NLL corrections.
We need a full NLL analysis to make a definite conclusion.
%------------------ Acknowledgment --------------------

This work was supported in part by the Monbusho
Grant-in-Aid Scientific Research No. A (1) 08304024,
No. C (1) 09640364.
%------------------ References -------------------------
%\vspace{-5mm}

\end{document}